%% file: 20091224_PHOBOS_Triggered_Correlations_Paper.tex
\documentclass[aps,prl,twocolumn,showpacs,groupedaddress]{revtex4}  

\usepackage{graphicx}    
\usepackage{dcolumn}    
\usepackage{bm, amsmath, amsthm, amssymb}     
\usepackage[footnotesize,figbotcap]{subfigure} 

\begin{document}

\title{High transverse momentum triggered correlations over a large pseudorapidity acceptance in Au+Au collisions at $\sqrt{s_{_{NN}}}$=200 GeV }

\input phobos_authors.tex 

\date{\today}

\begin{abstract}
A measurement of two-particle correlations with a high transverse momentum trigger particle ($p_T^{trig} > 2.5$~GeV/c) is presented for Au+Au collisions at $\sqrt{s_{_{NN}}}$=200 GeV over the uniquely broad longitudinal acceptance of the PHOBOS detector ($-4 < \Delta\eta < 2$).  A broadening of the away-side azimuthal correlation compared to elementary collisions is observed at all $\Delta\eta$.  As in p+p collisions, the near-side is characterized by a peak of correlated partners at small angle relative to the trigger particle.  However, in central Au+Au collisions an additional correlation extended in $\Delta\eta$ and known as the `ridge' is found to reach at least $|\Delta\eta| \approx 4$.  The ridge yield is largely independent of $\Delta\eta$ over the measured range, and it decreases towards more peripheral collisions.  For the chosen $p_T^{trig}$ cut, the ridge yield is consistent with zero for events with less than roughly 100 participating nucleons.
\end{abstract}


\maketitle 

One of the most fundamental discoveries at the Relativistic Heavy Ion Collider (RHIC) is that hard-scattered partons interact strongly as they traverse the produced medium \cite{PHOBOS_WhitePaper}.  In central Au+Au collisions, the suppression of high $p_T$ particle yields \cite{PHOBOS_RAA,PHENIX_RAA,STAR_RAA} and the disappearance of back-to-back high $p_T$ pairs \cite{STAR_BackToBackJets} are consistent with surface emission of jets due to the presence of an opaque medium that completely absorbs jets produced within or directed towards the interior.  

However, the energy and momentum of the away-side jet must be present in the final state, motivating the study of correlations between high $p_T$ trigger particles and lower $p_T$ associated particles.  Previous such measurements have shown that the absorbed jet energy manifests itself in an away-side structure centered at \mbox{$\Delta\phi=\pi$} that is broad in $\Delta\phi$ compared to p+p \cite{PHENIX_LowPtAssoc,STAR_LowPtAssoc} and in an enhanced correlation near $\Delta\phi\approx0$ observed by the STAR detector to extend at least to \mbox{$|\Delta\eta|<1.7$}, a phenomenon called the `ridge' \cite{STAR_LowPtAssoc,STAR_Ridge_Preprint,STAR_LongRange}.  
Although the ridge has been qualitatively described in this narrow $\Delta\eta$ region by a diverse assortment of proposed mechanisms \cite{Armesto,Hwa,Shuryak,Pantuev,Wong,Gavin}, the origin of the structure is still not well understood.  

The correlated yield of charged hadrons associated with a high transverse momentum trigger particle (\mbox{$p_T > 2.5$}~GeV/c) has been measured in collisions of gold nuclei at $\sqrt{s_{_{NN}}}$=200 GeV by the PHOBOS experiment at RHIC.  The data are presented as a function of pseudorapidity ($\Delta\eta$) and azimuthal angle ($\Delta\phi$) relative to the trigger particle over a range of collision centralities, where the number of participating nucleons ($N_{part}$) has been determined as in previous papers~\cite{PHOBOS_RAA,PHOBOS_WhitePaper}.   This measurement allows a unique examination of the ridge (and its dependence on event centrality) at large relative pseudorapidity up to $|\Delta\eta|\approx4$, placing a quantitative constraint on the possible interpretations of particle production correlated with high $p_T$ trigger particles.  

The PHOBOS detector \cite{PHOBOS_NIM} has two azimuthally opposed spectrometer arms, each covering 0.2 radians, which are used to select charged trigger tracks with \mbox{$p_T>2.5$}~GeV/c within an acceptance of \mbox{$0<\eta^{trig}<1.5$}.  Associated particles that escape the beam pipe (\mbox{$p_T$  
\scriptsize$\gtrsim$
\normalsize $7$~MeV/c at $\eta=3$, $p_T$ 
\scriptsize$\gtrsim$ 
\normalsize $35$}~MeV/c at $\eta=0$) are detected in a single layer of silicon comprising the octagon subdetector ($|\eta|<3$).  The loss of acceptance due to gaps in the octagon array is partly recouped by using the first layers of the vertex and spectrometer detectors.  To account for occupancy effects in the octagon and vertex detectors, each associated hit is weighted by the local occupancy (number of charged particles
per hit pad), estimated as in \cite{PHOBOS_Flow_vs_Energy} from the fraction of hit pads in the surrounding region.

The construction of the per-trigger correlated yield of charged particles
\large $\frac{1}{N_{trig}} \frac{d^{2}N_{ch}}{(d\Delta\phi) (d\Delta\eta)}$
\normalsize is described in any given centrality class by the following expression:

\begin{equation*}
B(\Delta\eta)  \cdot \left[\frac{s(\Delta\phi,\Delta\eta)}{b(\Delta\phi,\Delta\eta)}  - a(\Delta\eta) [ 1+2V(\Delta\eta) cos(2\Delta\phi) ]\right],
\label{eqn:CorrelatedYield}
\end{equation*}

\noindent where $N_{ch}$ is the number of charged particles and $N_{trig}$ the numbers of trigger particles.
The raw per-trigger distribution of same-event pairs $s(\Delta\phi,\Delta\eta)$ is divided by the raw mixed-event background distribution $b(\Delta\phi,\Delta\eta)$ to account for random coincidences and acceptance effects.  This ratio is calculated in $1$~mm bins of vertex position along the beam and averaged over the range $-15<z<10$~cm.  
$B(\Delta\eta)$, which converts the flow-subtracted correlation into a conditional yield, is constructed from the single-particle distribution ($dN/d\eta$) \cite{dNdeta} by convolution with the normalized $\eta$ distribution of trigger particles. 

Because `elliptic flow' -- the $[1+2v_{2}cos(2(\phi-\Psi_{RP}))]$ distribution of particles with respect to the impact parameter vector ($\Psi_{RP}$) -- is erased in the mixing of tracks and hits from different events, the remaining $\langle v_2^{trig} v_2^{assoc}\rangle$ modulation carried by $s(\Delta\phi,\Delta\eta)$ must also be subtracted.  It is approximated by the product \mbox{$V(\Delta\eta) = \langle v_{2}^{trig} \rangle \langle v_{2}^{assoc} \rangle$} using an event plane method that is sensitive to the RMS of $v_2$ rather than the mean \cite{PHOBOS_eccfluc}.  As such, this approximation is robust against $v_2$ fluctuations from initial source eccentricity \cite{PHOBOS_v2fluc_PRL} that could otherwise introduce up to a 10\% difference.  The flow magnitudes are calculated from a parameterization based on published PHOBOS measurements of $v_2$ as a function of  $N_{part}$, $p_T$, and $\eta$ \cite{PHOBOS_Flow}, assuming a factorized form.   The $v_2$ of the trigger track is corrected for occupancy effects in the spectrometer; the $v_2$ of associated hits is corrected for secondaries.  Both of these effects tend to suppress the magnitude of the observed $v_2$.  

The scale factor $a(\Delta\eta)$ is introduced to account for the small difference in multiplicity between signal and mixed-event distributions.  Its value -- within a few percent of unity in all cases considered -- is extracted separately in bins of centrality and $\Delta\eta$, using an implementation of the zero yield at minimum (ZYAM) method \cite{ZYAM} that fits the minimum of 
\large $\frac{s(\Delta\phi)/b(\Delta\phi)}{1+2Vcos(2\Delta\phi)}$ 
\normalsize with a second-order polynomial in the region $0.5<\Delta\phi<1.5$~radians.  

The hypothesis of zero correlated yield at the minimum in $\Delta\phi$ can be explored quantitatively from the evolution of $a(\Delta\eta)$, parameterized at each centrality as the sum of a Gaussian in $\Delta\eta$ and a term independent of $\Delta\eta$~\cite{Stephans_QM09_Proceedings}.
The deviation from \mbox{$a(\Delta\eta)=1$} or `absolute normalization' \cite{PHENIX_AbsNorm} can be characterized completely by the following two effects.
First, a bias of the $p_T$-triggered signal distribution to higher multiplicity accounts for the constant term in $a(\Delta\eta)$, the magnitude and centrality dependence of which are consistent with a prediction based on the centrality bin widths used in the analysis.  
Second, the presence of `cluster-like' particle production is reflected in the Gaussian term of $a(\Delta\eta)$, whose width is consistent with that extracted from inclusive two-particle correlations~\cite{PHOBOS_clusters}; the centrality dependence of the Gaussian amplitude is also in agreement with that found in~\cite{PHOBOS_clusters} after accounting for the dominant $1/N_{ch}$ dependence due to a trivial dilution by uncorrelated particles.  

\begin{figure}[htbp]
\centering
	   \includegraphics[width=0.45\textwidth, viewport=20 10 550 300]{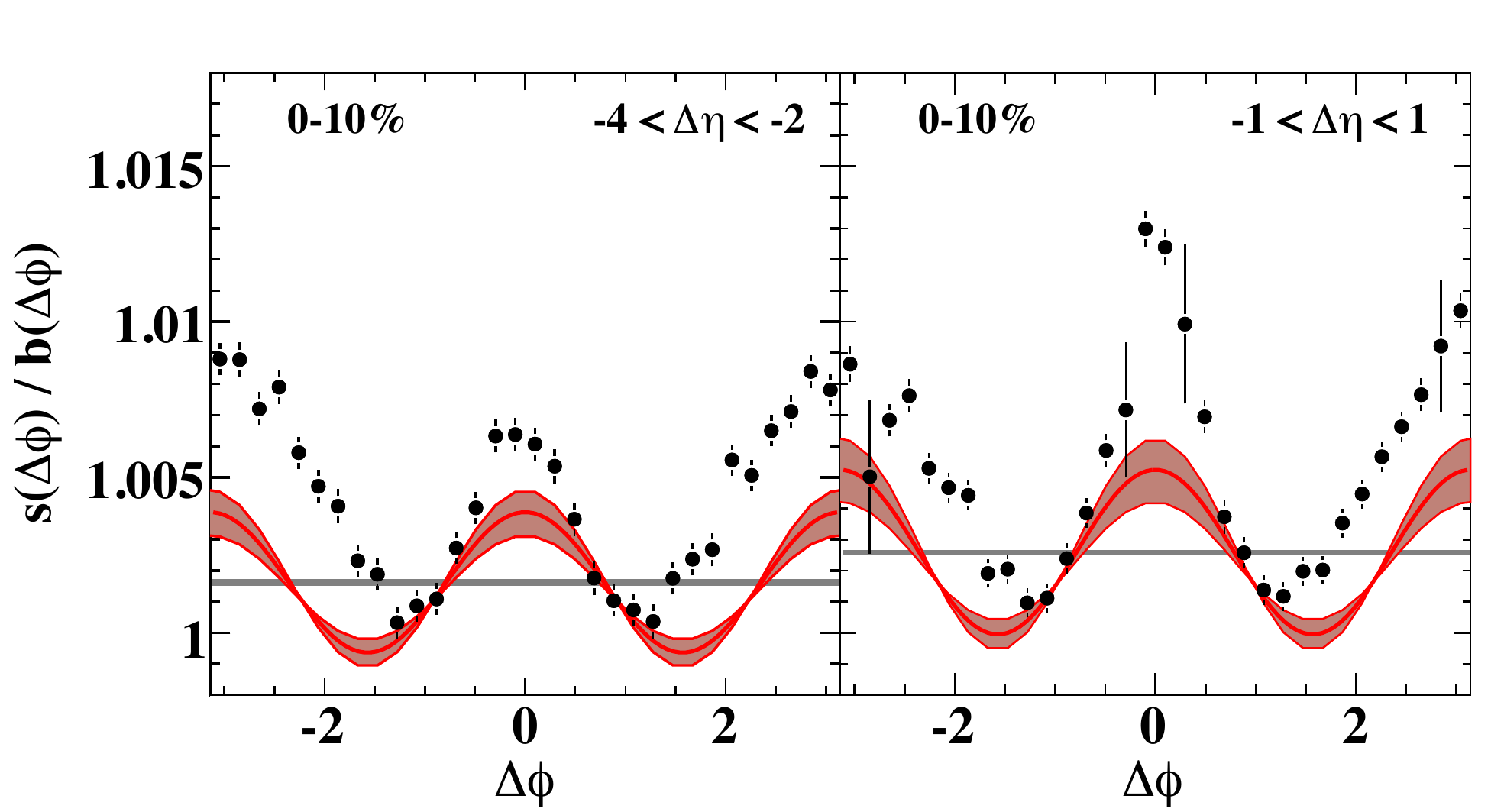}
\caption{(color online) Ratio of signal to background for the 10\% most central collisions at long-range (left) and short-range (right).  The estimated flow modulation (scaled by $a(\Delta\eta)$) and its uncertainty are represented by the solid line and shaded band.  The uncertainty on the ZYAM parameter is represented by the width of the narrow horizontal line.}
\label{fig:flowsubtraction}
\end{figure}

The uncertainty on the value of $a(\Delta\eta)$ is estimated from the error matrix of a 2-D fit to the extracted ZYAM parameters in $N_{part}$ and $\Delta\eta$.
The subtraction of elliptic flow from the raw correlation using this implementation of the ZYAM method is illustrated in Fig.~\ref{fig:flowsubtraction}, where the shaded band represents the flow uncertainty and the horizontal line the extracted ZYAM scale factor, $a(\Delta\eta)$. 

The dominant systematic error in this analysis comes from the uncertainty in estimating the magnitude of $\langle v_2^{trig}\rangle\langle v_2^{assoc}\rangle$.  This flow uncertainty is typically on the order of 15-20\%.  It reaches 50\% in the most central 3\% of collisions, although here $V(\Delta\eta)$ is quite small ($\leq 0.0008$) compared to the resulting jet correlation.

\begin{figure}[htbp]
\centering
   \subfigure[~p+p PYTHIA (version 6.325)]{
   	\includegraphics*[width=.4\textwidth, viewport=20 10 550 360]{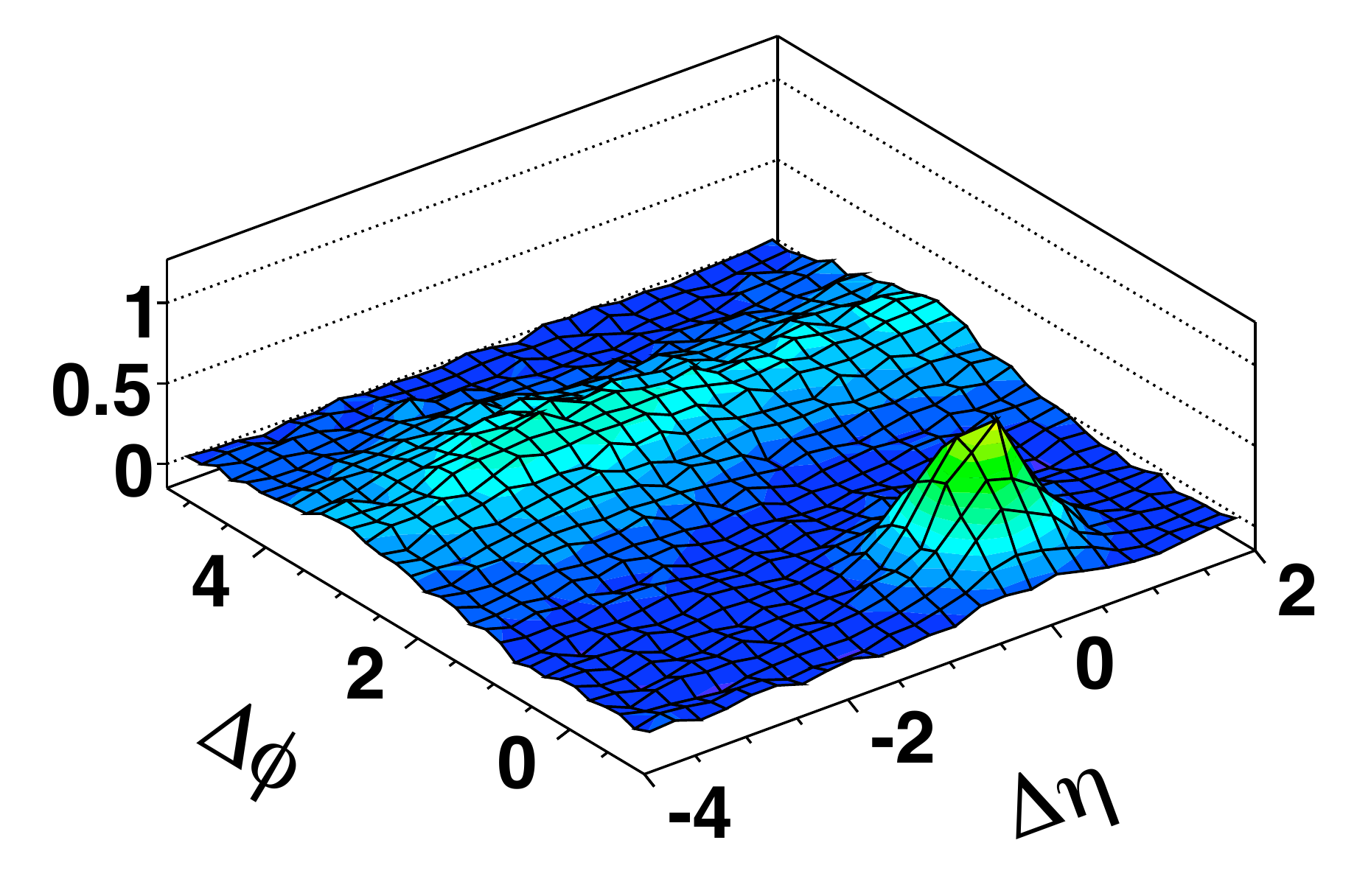} 
	\label{fig:pythiacorr}
   }
   \subfigure[~Au+Au 0-30\% (PHOBOS)]{
   	\includegraphics*[width=.4\textwidth, viewport=20 10 550 360]{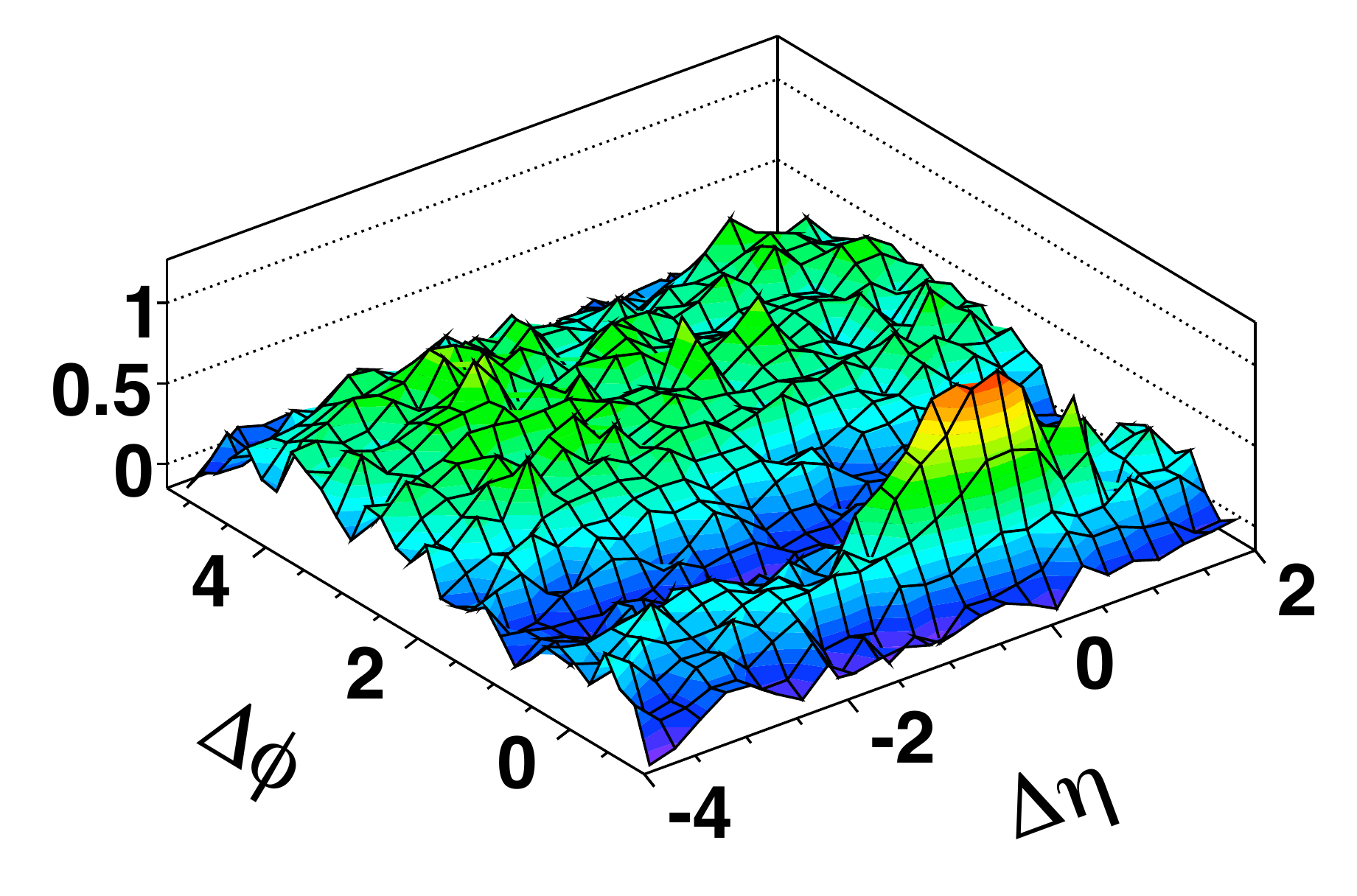} 
	\label{fig:auaucorr}
   }
   \subfigure[~Near-side $\Delta\eta$ projection ($|\Delta\phi| < 1$)]{
   	\includegraphics*[width=.45\textwidth]{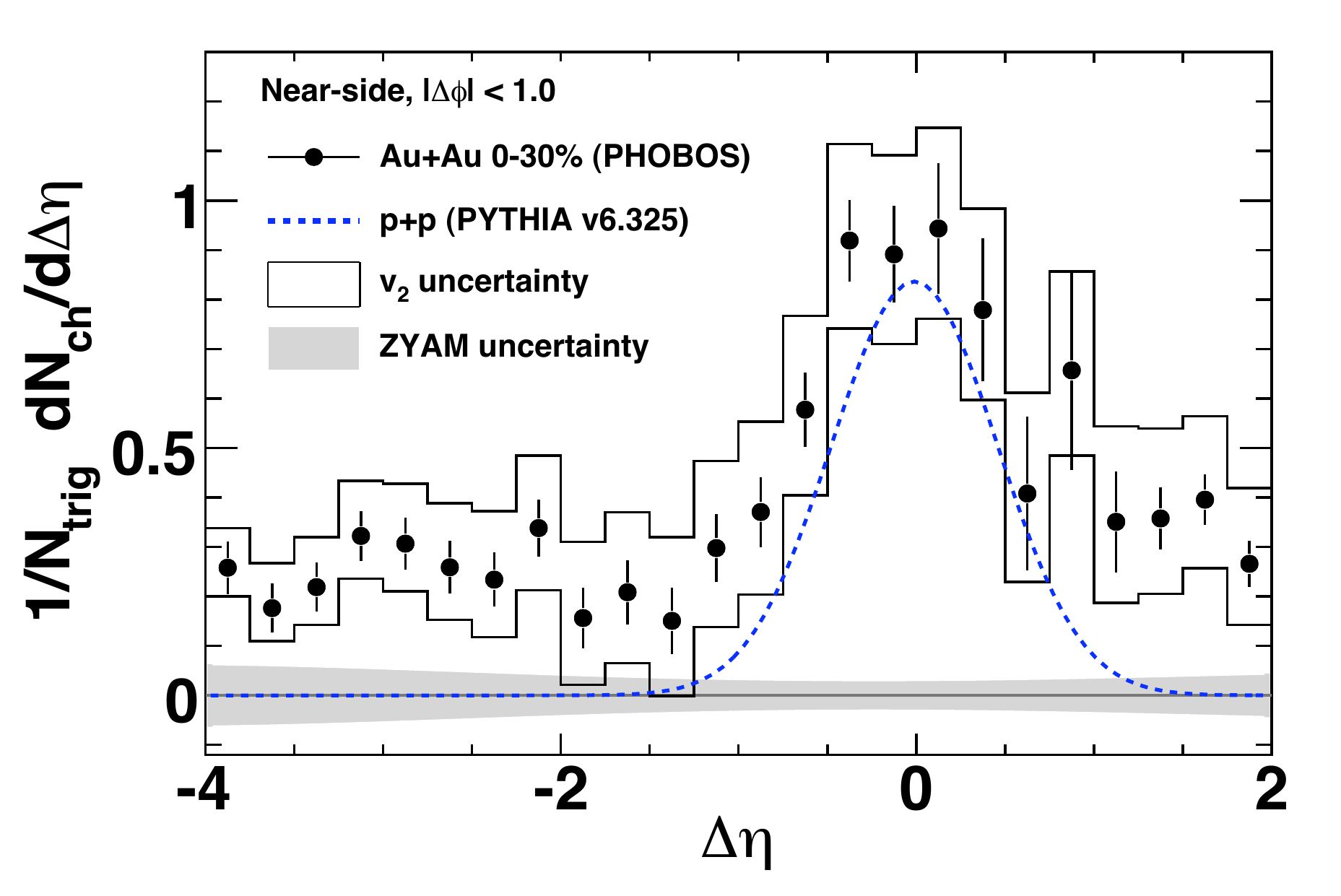}
         \label{fig:detaproj}	
   }
\caption {
(color online) Per-trigger correlated yield with \mbox{$p_T^{trig} > 2.5$}~GeV/c as a function of $\Delta\eta$ and $\Delta\phi$ for  
 $\sqrt{s}$ and $\sqrt{s_{_{NN}}}$=200 GeV \subref{fig:pythiacorr} PYTHIA p+p and \subref{fig:auaucorr} PHOBOS \mbox{0-30\%} central Au+Au collisions.  \subref{fig:detaproj} Near-side yield integrated over \mbox{$|\Delta\phi| < 1$} for 0-30\% Au+Au compared to PYTHIA p+p (dashed line) as a function of $\Delta\eta$.  Bands around the data points represent the uncertainty from flow subtraction. The error on the ZYAM procedure is shown as a gray band at zero. All systematic uncertainties are 90\% confidence level.
}
\label{fig:corrsurf}
\end{figure}

To understand the effects of the hot, dense medium on correlated particle production, the PHOBOS Au+Au data is compared to p+p events simulated with PYTHIA version 6.325 using default settings \cite{Pythia}.  
The $\Delta\eta$ and $\Delta\phi$ correlation structures in PYTHIA are compatible with published STAR p+p correlations \cite{STAR_LowPtAssoc} in the measured range for comparable analysis cuts. 
Except for the absence of elliptic flow subtraction, the analysis of PYTHIA simulations was identical to that for the Au+Au data.  The prominent features of the p+p correlation, shown in Fig.~\ref{fig:pythiacorr}, are a jet-fragmentation peak centered at $\Delta\phi=\Delta\eta=0$ and an away-side structure centered at $\Delta\phi = \pi$ that is similarly narrow in $\Delta\phi$ but extended in $\Delta\eta$, since the hard scattering can involve partons with very different fractions of the proton momentum.

\begin{figure*}[htbp]
\centering
	   \includegraphics[width=0.98\textwidth, viewport=5 5 570 220]{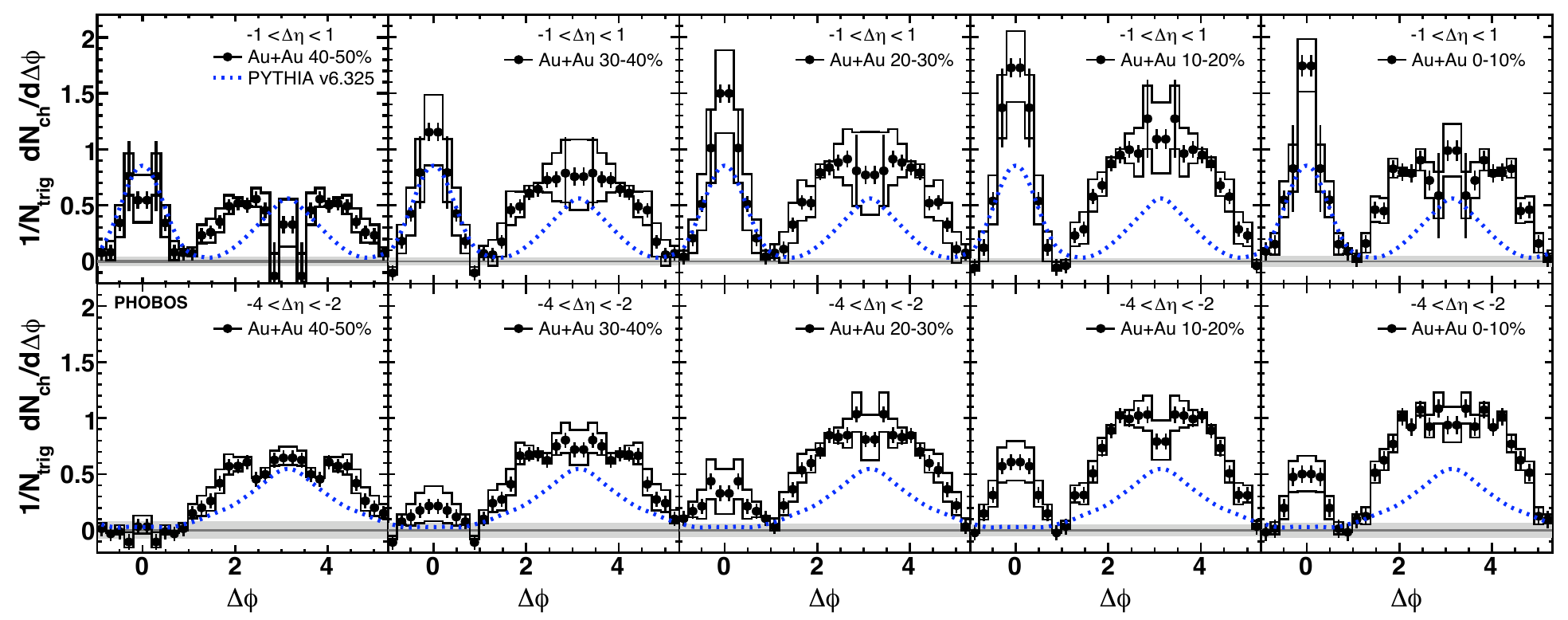} 
\caption{(color online) Projections of the correlated yield versus $\Delta\phi$ at short-range (top row, $-1<\Delta\eta<1$) and long-range (bottom row, $-4<\Delta\eta<-2$) for five centrality bins (most central on right).  Points have been reflected about $\Delta\phi=0$ and averaged.  The dashed line is p+p PYTHIA for comparison.  90\% C.L. systematic uncertainties are presented as in Fig 2c.}
\label{fig:dphiproj}
\end{figure*}

In central Au+Au collisions, particle production correlated with a high $p_T$ trigger particle is strongly modified as shown in Fig.~\ref{fig:auaucorr}.  Not only is the enhanced away-side yield much broader in $\Delta\phi$, the near-side peak at \mbox{$\Delta\phi\approx0$} now sits atop a pronounced ridge of correlated partners extending continuously and undiminished all the way to $|\Delta\eta|=4$.  
To examine the near-side structure more closely, the correlated yield is integrated over the region $|\Delta\phi|<1$ and plotted as a function of $\Delta\eta$ in Fig.~\ref{fig:detaproj}.  For the most central 30\% of Au+Au collisions, there is a significant and relatively flat correlated yield of about 0.25 particles per unit pseudorapidity far from the trigger particle.  

A more detailed examination of the correlation structure is possible by projecting the correlation onto the $\Delta\phi$ axis as in Fig.~\ref{fig:dphiproj}.  In the top row of that figure, the correlated yield in Au+Au is compared for five centrality bins (40-50\%, 30-40\%, 20-30\%, 10-20\% and 0-10\%) to PYTHIA-simulated p+p events at short-range (i.e. integrated over the region $|\Delta\eta|<1$).  In the bottom row, the same comparison is shown at long-range (i.e. integrated over the region $-4<\Delta\eta<-2$).  

\begin{figure}[htbp]
\centering
	   \includegraphics[width=0.5\textwidth, viewport=20 20 550 420]{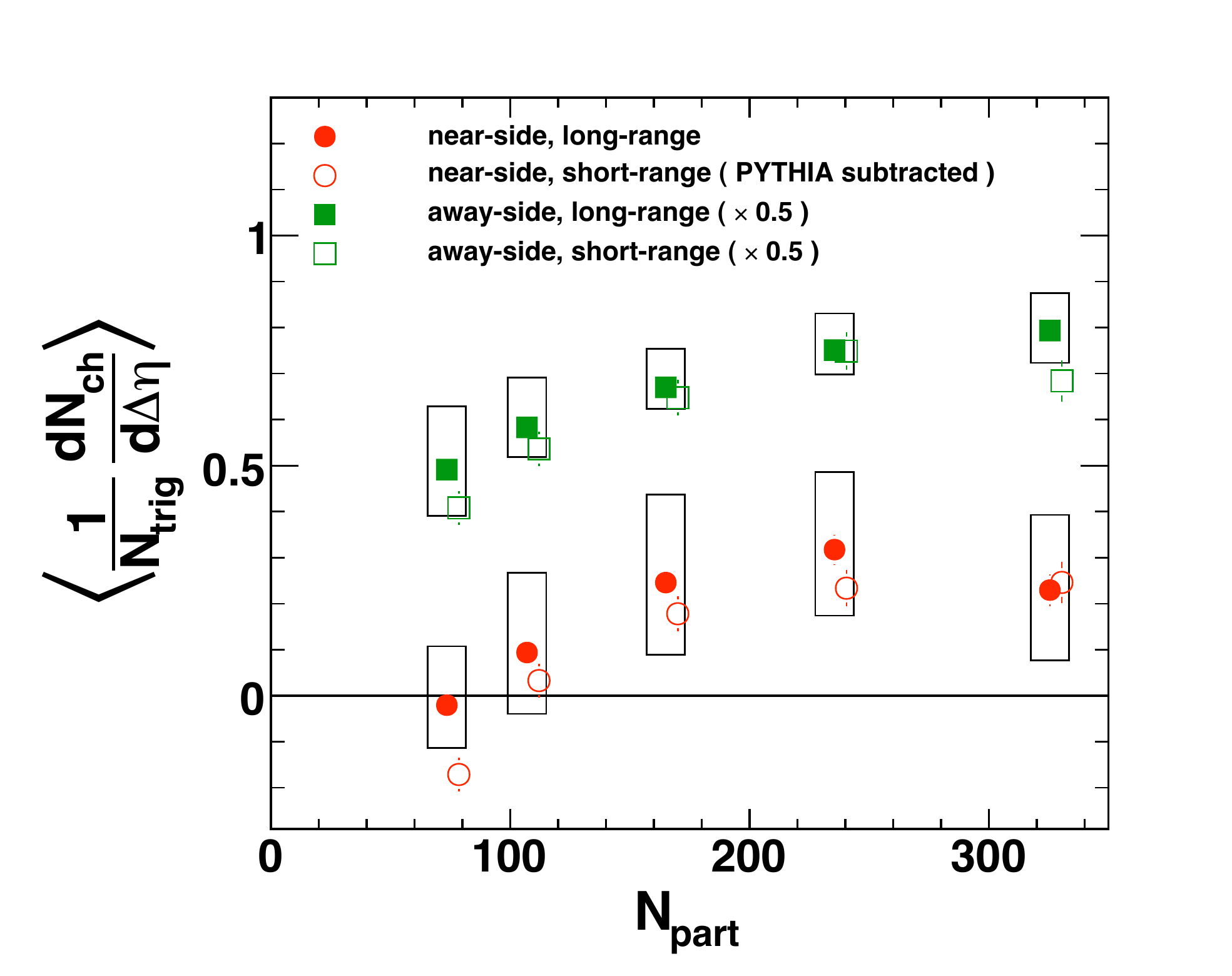} 
\caption{(color online) Average near-side ($|\Delta\phi|<1$) and away-side ($\Delta\phi>1$) yields as a function of $N_{part}$ at both short-range ($| \Delta\eta| < 1$) and long-range ($-4 < \Delta\eta < -2$).  The near-side yield at short-range has the full PYTHIA jet yield subtracted from it.  For clarity of presentation, the open points have been offset slightly and the away-side yields scaled down by a factor of 2.  Boxes correspond to the combined 90\% systematic uncertainties on the $v_2$ estimate and ZYAM procedure.}
\label{fig:RidgeVsNpart}
\end{figure}

Focusing first on the away-side correlation, a number of features become apparent.  First, the shape of the correlation is considerably broader in $\Delta\phi$ for Au+Au collisions compared to p+p in all measured centrality bins.  Additionally, the magnitude of the away-side yield is enhanced relative to p+p, increasingly so for more central Au+Au collisions.  Finally, the away-side correlation seems to have a very similar shape and centrality dependence at both short- and long-range.  This last observation is explored more quantitatively in Fig.~\ref{fig:RidgeVsNpart}, where integrated away-side yields ($\Delta\phi>1$) are presented as a function of participating nucleons at short- and long-range.

The near-side region also shows a strong modification spanning the full measured pseudorapidity range.  At short-range (top row of Fig.~\ref{fig:dphiproj}), a narrow peak at $\Delta\phi \approx 0$ is observed.  In central collisions, this peak has a large contribution in excess of the p+p jet yield.  The near-side, short-range correlation decreases in magnitude with decreasing centrality, reaching the same height as p+p in the 40-50\% bin.  

At long-range (bottom row of Fig.~\ref{fig:dphiproj}), the persistence of the ridge correlation to very large $\Delta\eta$ is evident in the peak at $\Delta\phi \approx 0$ for central Au+Au collisions.  This effect is completely absent in more elementary systems.  The ridge yield also decreases in magnitude for more peripheral collisions until it disappears in the 40-50\% bin.  

The similar centrality dependence of the short- and long-range yields in excess of the p+p jet correlation suggests a decomposition of the near-side correlation into distinct jet and ridge components.  Such a separation is supported by previous STAR measurements of the associated particle $p_T$ spectra, the centrality independence of the jet-like yield, and the ridge-subtracted fragmentation function \cite{STAR_Ridge_Preprint}.  

In Fig.~\ref{fig:RidgeVsNpart}, the integrated ridge yield at long-range (filled circles) is compared to the short-range yield (open circles) after subtraction of a jet component corresponding to the PYTHIA yield.
A more precise agreement between these two yields might be achieved for an approximately $15\%$ smaller jet component, inducing all the open circles to move up by $0.07$.
Nonetheless, already for this simple assumption, the ridge yield is shown to be the same within experimental uncertainties at all $\Delta\eta$.  It decreases as one goes towards more peripheral collisions and is consistent with zero in the most peripheral bin analyzed (40-50\%).  While the systematic errors do not completely exclude a smooth disappearance of the ridge as one approaches p+p collisions, these data suggest that with a 2.5 GeV/c trigger particle the ridge may already have disappeared by $N_{part}=80$ (about 45\% central).

In this letter, the long-range correlation of particles aligned azimuthally with a high $p_{T}$ trigger particle -- a unique feature of heavy ion collisions -- is shown to extend over at least four units of pseudorapidity.
This represents a dramatic extension of previous measurements, providing powerful constraints on the nature of particle production in heavy ion collisions.
The persistence of this correlation to such large pseudorapidity separations suggests it is imprinted in the earliest moments after the collision, naturally favoring interpretations that invoke global features of the system like collision geometry and radial flow, rather than local features like medium-modified fragmentation.


\input phobos_acknowledge.tex   

\bibliographystyle{apsrev}
\bibliography{prlbibfile}

\end{document}

%% file: phobos_authors.tex
%
\author{
B.Alver$^4$,
B.B.Back$^1$,
M.D.Baker$^2$,
M.Ballintijn$^4$,
D.S.Barton$^2$,
R.R.Betts$^6$,
A.A.Bickley$^7$,
R.Bindel$^7$,
W.Busza$^4$,
A.Carroll$^2$,
Z.Chai$^2$,
V.Chetluru$^6$,
M.P.Decowski$^4$,
E.Garc\'{\i}a$^6$,
T.Gburek$^3$,
N.George$^2$,
K.Gulbrandsen$^4$,
C.Halliwell$^6$,
J.Hamblen$^8$,
M.Hauer$^2$,
C.Henderson$^4$,
D.J.Hofman$^6$,
R.S.Hollis$^6$,
R.Ho\l y\'{n}ski$^3$,
B.Holzman$^2$,
A.Iordanova$^6$,
E.Johnson$^8$,
J.L.Kane$^4$,
N.Khan$^8$,
P.Kulinich$^4$,
C.M.Kuo$^5$,
W.Li$^4$,
W.T.Lin$^5$,
C.Loizides$^4$,
S.Manly$^8$,
A.C.Mignerey$^7$,
R.Nouicer$^{2,6}$,
A.Olszewski$^3$,
R.Pak$^2$,
C.Reed$^4$,
C.Roland$^4$,
G.Roland$^4$,
J.Sagerer$^6$,
H.Seals$^2$,
I.Sedykh$^2$,
C.E.Smith$^6$,
M.A.Stankiewicz$^2$,
P.Steinberg$^2$,
G.S.F.Stephans$^4$,
A.Sukhanov$^2$,
M.B.Tonjes$^7$,
A.Trzupek$^3$,
C.Vale$^4$,
G.J.van~Nieuwenhuizen$^4$,
S.S.Vaurynovich$^4$,
R.Verdier$^4$,
G.I.Veres$^4$,
P.Walters$^8$,
E.Wenger$^4$,
F.L.H.Wolfs$^8$,
B.Wosiek$^3$,
K.Wo\'{z}niak$^3$,
B.Wys\l ouch$^4$\\
\vspace{3mm}
\small
%
%
%
%
$^1$~Argonne National Laboratory, Argonne, IL 60439-4843, USA\\
$^2$~Brookhaven National Laboratory, Upton, NY 11973-5000, USA\\
$^3$~Institute of Nuclear Physics PAN, Krak\'{o}w, Poland\\
$^4$~Massachusetts Institute of Technology, Cambridge, MA 02139-4307, USA\\
$^5$~National Central University, Chung-Li, Taiwan\\
$^6$~University of Illinois at Chicago, Chicago, IL 60607-7059, USA\\
$^7$~University of Maryland, College Park, MD 20742, USA\\
$^8$~University of Rochester, Rochester, NY 14627, USA\\
}

%% file: phobos_acknowledge.tex
%
%
%
%
This work was partially supported by U.S. DOE grants 
DE-AC02-98CH10886,
DE-FG02-93ER40802, 
DE-FG02-94ER40818,  
DE-FG02-94ER40865, 
DE-FG02-99ER41099, and
DE-AC02-06CH11357, by U.S. 
NSF grants 9603486, 
0072204,            
and 0245011,        
by Polish MNiSW grant N N202 282234 (2008-2010),
by NSC of Taiwan Contract NSC 89-2112-M-008-024, and
by Hungarian OTKA grant (F 049823).